\documentclass[twocolumn,showpacs,preprintnumbers,prl,epsfig,superscriptaddress]{revtex4}

\usepackage{graphicx}
\usepackage{dcolumn}
\usepackage{bm}

\usepackage{array}  
\usepackage{multirow}  
\usepackage{textcomp} 

\begin{document}

\preprint{K. W. Kim: 2-col / Sr$_{2}$CuO$_{3}$}

\title{Bound Excitons in Sr$_{2}$CuO$_{3}$}

\author{K. W. Kim}
\email{kyungwan.kim@gmail.com}

\affiliation{Department of Physics, University of Fribourg, Chemin
du Mus$\acute{e}$e 3, CH-1700 Fribourg, Switzerland}

\affiliation{Condensed Matter Physics and Materials Science
Department, Brookhaven National Laboratory, Upton, New York 11973,
USA}

\affiliation{School of Physics and Research Center for Oxide
Electronics, Seoul National University, Seoul 151-747, Korea}

\author{G. D. Gu}
\affiliation{Condensed Matter Physics and Materials Science
Department, Brookhaven National Laboratory, Upton, New York 11973,
USA}

\author{C. C. Homes}
\affiliation{Condensed Matter Physics and Materials Science
Department, Brookhaven National Laboratory, Upton, New York 11973,
USA}

\author{T. W. Noh}
\affiliation{School of Physics and Research Center for Oxide
Electronics, Seoul National University, Seoul 151-747, Korea}

\date{\today}

\begin{abstract}
We investigated temperature dependent optical spectra of the
one-dimensional chain compound Sr$_{2}$CuO$_{3}$. The charge
transfer transition polarized along the chain direction shows a
strongly asymmetric line shape as expected in one-dimensional
extended Hubbard model. At low temperature the charge transfer
peak shows a large blue shift and reveals additional sharp peaks
at the gap. Even though many spectroscopic studies suggest that
this material can not have a bound exciton based on the
one-dimensional extended Hubbard model, we attribute the
additional sharp peaks to excitons, which come to exist due to the
long range Coulomb interaction.
\end{abstract}

\pacs{71.27.+a, 71.35.-y, 78.20.-e, 78.67.-n}

\maketitle

The physics of Mott insulators is still a subject of great
interest due to a possible link with high temperature cuprate
superconductors and its own physical properties. One-dimensional
(1D) Mott insulators in particular have attracted lots of
attention not only from unique physical properties such as
spin-charge separation but also as a good candidate for an optical
switching material thanks to their large optical nonlinearity
\cite{Nonlinearity-Nature, M.Ono, M.Ono-PRL, film-absorption, T.
Kidd}. Ni-halogen bridged 1D materials and 1D cuprates of
Sr$_{2}$CuO$_{3}$ and Ca$_{2}$CuO$_{3}$ show comparable or even
larger optical nonlinear property than well known nonlinear
materials such as conjugated polymers \cite{Nonlinearity-Nature,
M.Ono}. In those materials the optical gap has charge transfer
character. That is, the optical gap is determined by the occupied
halogen (oxygen) $p$-orbitals and empty Ni (Cu) $d$-orbitals. But
they should be metallic without strong onsite Coulomb repulsion.
Their nonlinear behavior has been explained with one-dimensional
extended Hubbard model (1D EHM), of which Hamiltonian is

\begin{equation}
H=-t\sum_{l,\sigma}(c^{\dag}_{l+1,\sigma}c_{l,\sigma}+h.c.)+
U\sum_{l}{n_{l,\uparrow}n_{l,\downarrow}} +V\sum_{l}{n_{l}n_{l+1}}
\label{eq:1D EHM}
\end{equation}

\noindent where $c^{\dag}_{l,\sigma} (c_{l,\sigma})$ is the
creation (annihilation) operator for a spin
$\sigma=\uparrow,\downarrow$ electron at site $l$,
$n_{l,\sigma}=c^{\dag}_{l,\sigma}c_{l,\sigma}$,
$n_{l}=n_{l,\uparrow}+n_{l,\downarrow}$, $t$ is the hopping
integral between nearest neighbor sites, and $U$ ($V$) is the
onsite (nearest neighbor) Coulomb interaction. In this model, a
bound exciton is formed when $V>2t$ \cite{DDMRG}. Among them,
Ni-Br-Br compound and Sr$_{2}$CuO$_{3}$ are believed to be on the
verge of forming a bound exciton which should affect their
nonlinear properties \cite{M.Ono}. For Ni-Br-Br compound, it is
clearly demonstrated by the low temperature measurements that
there is a bound exciton \cite{M.Ono-PRL}. However, for
Sr$_{2}$CuO$_{3}$ no clear evidence for a bound exciton has been
provided yet.

Sr$_{2}$CuO$_{3}$ has one-dimensional chain with CuO$_{4}$
plaquettes, which is a basic building block of the high
temperature cuprate superconductors. It is believed that the
physical properties of this material can be explained with 1D EHM.
Ono \textit{et al}. have pursued thorough studies on this material
and from electro-reflectance and photoconductivity measurements
they argued that this material should have a bound exciton
\cite{M.Ono}. However, electron energy loss (EELS) spectra and
optical spectra have been well explained with parameters which
expect no bound exciton \cite{EELS1, EELS1-comment, DDMRG-SCO}.

In this letter, we show a clear evidence of bound excitons in
Sr$_{2}$CuO$_{3}$ from temperature dependent optical spectra,
which appears as sharp peaks. Even though the overall line shape
of optical spectra could be understood with 1D EHM, we notice that
those sharp features are beyond 1D EHM. We propose that the long
range Coulomb interaction should be taken into account as a
missing ingredient to explain those sharp features.

Single crystalline samples were grown using the traveling-solvent
floating zone method. Temperature dependent polarized reflectivity
spectra were carefully measured over a wide energy range. In a low
energy region of 30 to 24000 cm$^{-1}$ (4 meV to 3 eV) an in situ
evaporation technique was adopted in the overfilling method on
Bruker 66v /S Fourier transform spectrometer. In 4000-50000
cm$^{-1}$ (0.5 eV to about 6 eV) Cary5 grating spectrometer was
used in the underfilling method. High energy spectra in 6-30 eV
were measured at room temperature utilizing synchrotron radiation
from the normal incidence monochromater beam line at Pohang Light
Source (PLS). All measurements were done on freshly cleaved
surfaces. More careful measurements were performed in the visible
region where the strongly anisotropic charge transfer response is.
For a better polarization, two polarizers (a polarizer-analyzer
configuration) were used. The polarization for measurements was
kept close to \textit{s}-polarization to minimize possible
rotation of polarization after reflections on the sample and
mirrors. And those angles of incidence from the sample and mirrors
between two polarizers were less than 10 degrees. The complex
optical conductivity spectra
$\tilde{\sigma}(\omega)$ 
were obtained from Kramers-Kronig transformation of reflectivity
$\emph{R}(\omega)$.

\begin{figure}
\includegraphics[width=8cm]{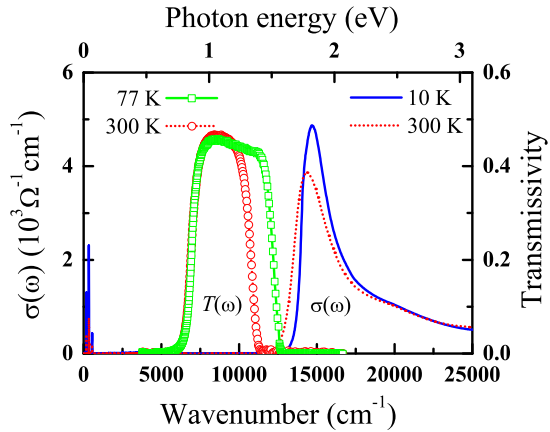}
\caption{Optical conductivity and transmission spectra along the
chain direction of Sr$_{2}$CuO$_{3}$. Transmission was measured on
a piece cleaved to be about 200 $\mu$m thick.}
\label{fig.cond-tra}
\end{figure}

\begin{figure}
\includegraphics[width=8cm]{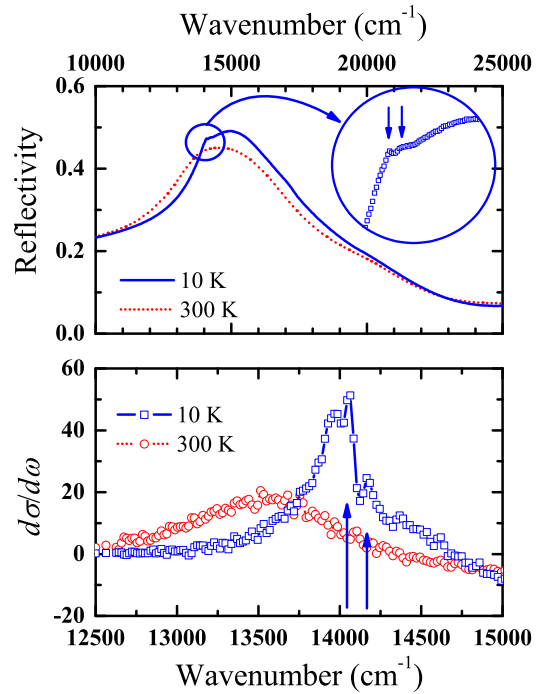}
\caption{Reflectivity and derivative spectra of $\sigma(\omega)$
along the chain direction. Sharp peaks at 10 K are marked with
arrows. The reflectivity spectra were measured by 1 nm resolution
and the instrumental error is smaller than the line thickness
(symbol size) in this energy region.} \label{fig.ref-dcond}
\end{figure}

\begin{figure}
\includegraphics[width=8cm]{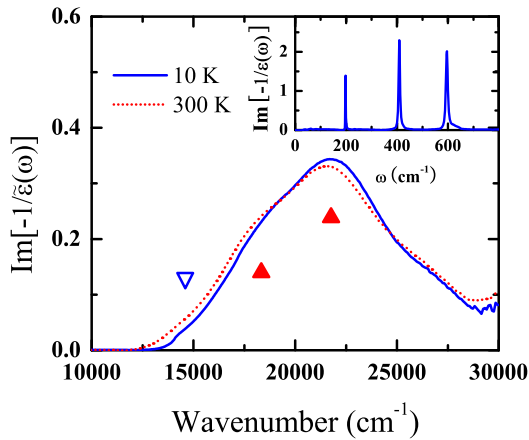}
\caption{Loss function of Sr$_{2}$CuO$_{3}$ along the chain
direction. The solid triangles indicate peaks observed in the EELS
measurement and the open triangle indicates the excitonic response
at 10 K. The inset shows a 10 K spectrum in phonon region, peaks
of which correspond to LO phonon modes.} \label{fig.loss-ftn}
\end{figure}

Because of interesting properties of this material as described
above, optical spectra have been already reported many times
\cite{M.Ono, Nonlinearity-Nature, film-absorption, T. Kidd}. They
agree with each other in general features that it has strong
asymmetric charge transfer peak at 1.5-2.5 eV along the chain as
shown again in fig.~\ref{fig.cond-tra} and no apparent feature
along other axes perpendicular to the chain in that energy region.
As temperature is lowered this peak moves to higher energy and
becomes narrower. But it should be noted that there is a distinct
difference between our spectra and reported data. Most of previous
reports show a long tail into the gap region. But as shown in
fig.~\ref{fig.cond-tra}, Sr$_{2}$CuO$_{3}$ is fairly transparent
in the gap region, which dictates that the optical conductivity
should not have that long tail into the gap region. Thanks to high
quality large single crystals and accurate measurements over a
wide energy range, we could get reasonable optical spectra with a
sharp gap edge.

This strong charge transfer peak can be well explained within 1D
EHM, parameters of which expect no bound exciton \cite{DDMRG,
EELS1}. The temperature dependent change of this peak has been
attributed to electron phonon coupling as in other insulating
cuprates \cite{M.Ono, La2CuO4, ARPES CCOC}. However, there is
another distinctive change at low temperature.
Figure~\ref{fig.ref-dcond} shows temperature dependent
$\emph{R}(\omega)$ and derivative spectra of $\sigma(\omega)$.
There are clear additional sharp features only at low temperature.
Note that the strong charge transfer peak which steeply increases
just above the gap overwhelms those features. As a result, these
sharp peaks in reflectivity seem to be hidden in $\sigma(\omega)$.
However, its derivative also clearly reveals their sharp nature.
The loss function in fig.~\ref{fig.loss-ftn} also shows that there
is a separated small feature (open triangle) below the strong
charge transfer peak \cite{K. W. Kim-SCO2}. These sharp peaks have
been also observed in a photoluminescence (PL) measurement with a
finite Stokes shift of about 600 cm$^{-1}$ \cite{M.Ono}. The
energy difference ($\sim$130 cm$^{-1}$) of the two peaks marked as
arrows in fig.~\ref{fig.ref-dcond} is same with that of peaks in
PL spectra. The inset of fig.~\ref{fig.loss-ftn} shows the loss
function by longitudinal optic (LO) phonons. It is clear that the
observed Stokes shift in PL spectra is by the highest LO phonon,
which corresponds to the Cu-O breathing mode.

Ono \textit{et al}. argued that this material has a bound exciton
from photoconductivity and electro-reflectance spectra
\cite{M.Ono}. However, EELS and $\sigma(\omega)$ require the
system not to have a bound exciton within 1D EHM \cite{EELS1,
DDMRG-SCO}. It should be noted that the observed sharp peaks are
different from the exciton in 1D EHM. The line shape of
$\sigma(\omega)$ in 1D EHM depends on the strength of the
intersite Coulomb interaction $V$. As $V$ increases, more spectral
weight shifts to lower energy and $\sigma(\omega)$ has a strongly
asymmetric shape like spectra shown fig.~\ref{fig.cond-tra}
\cite{DDMRG}. Note that any finite $V$ can result in a bound
exciton at zone boundary, which has been manifested in EELS
spectra of this material \cite{EELS1, long-range}. If $V>2t$, a
holon (a hole without spin) and a doublon (a double occupancy
site) form a bound exciton even at zone center and more spectral
weight is accumulated to the exciton, resulting in a strong narrow
peak just above the gap. Although there could be other peaks
corresponding to continuum excitation or multiple excitons, they
are much weaker and should come at higher energy than that of the
exciton. Therefore, according to 1D EHM with appropriate
parameters, there is only one strong peak expected at just above
the gap in Sr$_{2}$CuO$_{3}$ regardless of forming a bound
exciton. However, the sharp peaks observed here come right at the
gap and have much smaller weight compared to the main charge
transfer peak, which 
can not be explained by 1D EHM.

 Moskvin \textit{et al}. proposed a weak one-center transition
 localized within one CuO$_{4}$ plaquette should come at slightly
lower energy than the two-center transition between neighboring
plaquettes, which corresponds to the strong charge transfer peak
observed at 1.8 eV in fig.~\ref{fig.cond-tra} \cite{EELS2}. It was
argued that this one center peak should be observed both along the
chain and perpendicular to the chain directions, which correspond
to $b$ and $a$ axes respectively. Figure~\ref{fig.loss-ftn} shows
the loss function at zone center. There are clearly two peaks at
both temperatures marked with solid triangles which were observed
in EELS spectra. But both peaks turned out to have the two-center
transition character, and the small feature appears additionally
only at low temperature (open triangle) \cite{K. W. Kim-SCO2}.
Note that no feature has been observed in an absorption
measurement along $a$ direction in this energy region
\cite{film-absorption}. Therefore, the one-center transition can
not explain the observed sharp peaks \cite{K. W. Kim-SCO2}. As
another candidate for these peaks, onsite \textit{d-d} transitions
should be discussed. In many insulating cuprates, onsite
\textit{d-d} transitions have been observed in this energy region
\cite{CuB2O4, CuGeO3, SCOC}. In materials with CuO$_{2}$ planes,
these \textit{d-d} transitions are usually very weak.
Interestingly in CuB$_{2}$O$_{4}$ with 0 dimensional CuO$_{4}$
plaquette, these \textit{d-d} transitions appear as very sharp
peaks even at room temperature \cite{CuB2O4}. However, the
featureless absorption along $a$ axis also eliminates this
possibility for the sharp peaks. If there were onsite \textit{d-d}
transitions, they are expected to be stronger (sharper) in $a$
axis (0D CuO$_{4}$ plaquette) than in $b$ axis (1D chain of
CuO$_{3}$), which is contrary to the observations here. As being
discussed later, the observed strength of these peaks is even
stronger than that of 0D case of CuB$_{2}$O$_{4}$. Note that any
localized phenomena within a CuO$_{4}$ plaquette can not reconcile
the lack of absorption along $a$ axis of
Sr$_{2}$CuO$_{3}$. 

Even though these sharp peaks are not discussed carefully, it has
been noticed that there could be more than one transition in the
charge transfer peak in Sr$_{2}$CuO$_{3}$ \cite{M.Ono, el-ph}.
Matsueda \textit{et al}. proposed that a strong electron phonon
interaction should be considered to understand the charge transfer
peak in Sr$_{2}$CuO$_{3}$ \cite{el-ph}. They argued that the
effect of electron phonon interaction is enhanced by onsite
Coulomb interaction and it could result in splitting of the charge
transfer peak. Not only in this system but also in many insulating
cuprates, the electron phonon interaction could play an important
role \cite{ARPES CCOC, M.Ono}. 
However, the observed sharp peaks are qualitatively different from
what is expected due to the electron phonon interaction. Note that
the calculated spectra have always the strongest peak just above
the gap with expected parameters for Sr$_{2}$CuO$_{3}$ in 1D EHM
\cite{DDMRG-SCO, EELS1,film-absorption}. Another point which has
to be noticed is that the energy difference ($\sim$130 cm$^{-1}$)
of the two peaks can not be explained by any phonon observed in
this system \cite{Raman}.

Interestingly quite a similar behavior has been observed in a
Ni-Br-Br compound which is also supposed to be close to the
boundary of $V=2t$ within 1D EHM. Its reflectance at 77 K shows a
narrow peak on top of the strong charge transfer peak
\cite{M.Ono}, which looks similar to that of Sr$_{2}$CuO$_{3}$ at
10 K. Even though this multi-peak structure does not show up as
separated two peaks in $\epsilon_{2}$, it is clear that there has
to be more than one peak \cite{M.Ono, Nonlinearity-Nature}. The
spectrum at 4.2 K clearly shows a few peaks with most of the
spectral weight accumulated at the first peak, which is an exciton
\cite{M.Ono-PRL}. This similarity of two materials suggests that
such a small peak observed just above the gap could be a unique
phenomenon in an 1D system with $V \sim 2t$, which is close to the
boundary of forming a bound exciton in 1D EHM.

An exciton in semiconductor is a hydrogen-like bound state of
electron and hole due to the long range Coulomb interaction
\cite{book}. On the other hand, most theoretical studies of the 1D
EHM have usually considered onsite and nearest neighbor Coulomb
interactions instead of the long range interaction, which expect
one strong exciton at the gap. There have been reported a few
theoretical studies on 1D EHM with the long range Coulomb
interaction \cite{long-range, long-range2}. Interestingly, it is
agreed that even in strongly correlated systems with large $U$,
the long range interaction gives rise to hydrogen-like bound
states as in semiconductors. Such a Wannier exciton could be
formed with a smaller nearest neighbor interaction than the case
without the long range interaction. In 1D EHM of eq.~(\ref{eq:1D
EHM}) with $U\gg t$, the binding energy of a bound exciton is
$V-4t+4t^{2}/V$ and its size is $V^{2}/(V^{2}-4t^{2})$ (unit cell
lattice parameter=1) when $V>2t$ \cite{DDMRG}. It is clear that as
$V$ approaches $2t$ the size diverges, which is natural for an
exciton with small binding energy. This situation is still the
same in the case with a finite $U$ \cite{DDMRG}. In this limit of
$V\sim 2t$, there should be significant correlation between holon
and doublon at distance larger than one unit cell. Therefore the
long range Coulomb interaction could play an important role.
Although there is no consensus whether there is a bound exciton or
not, Sr$_{2}$CuO$_{3}$ must be closely located to the boundary of
$V=2t$ within the 1D EHM. Therefore, the observed sharp peaks at
low temperature could be responses of Wannier-like excitons with
help of the long range Coulomb interaction.

\begin{figure}
\includegraphics[width=8cm]{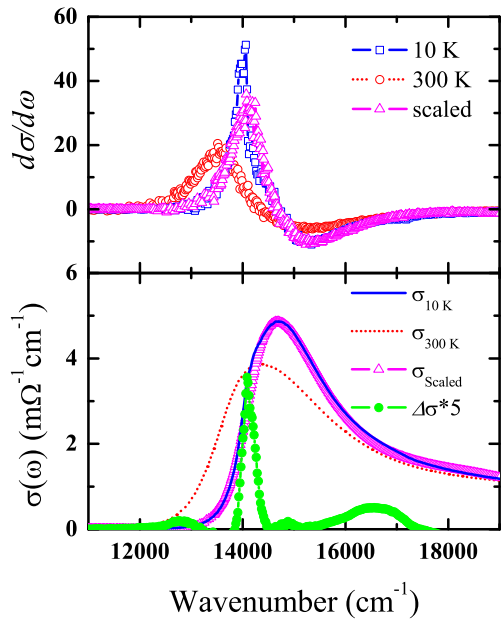}
\caption{Scaling of $d\sigma/d\omega$ and $\sigma(\omega)$ by peak
height and width. The 10 K spectra and the scaled 300 K spectra
agree with each other except the sharp peaks.} \label{fig.scaling}
\end{figure}

The optical spectra at 10 and 300 K look quite similar to each
other except the sharp peaks at 10 K. 
Figure~\ref{fig.scaling} shows $d\sigma/d\omega$ and
$\sigma(\omega)$ at 10 and 300 K and the scaled 300 K spectra
simply by peak height and width. The scaled spectra remarkably
agree with 10 K spectra except the sharp exciton peaks. As in
other insulating cuprates, such a temperature dependence could be
attributed to an electron phonon coupling \cite{M.Ono, La2CuO4,
ARPES CCOC}. Assuming the overall response modulated by
temperature and electron-phonon coupling could be counted by the
scaling, the excitonic response was estimatied from the difference
spectrum as shown in fig.~\ref{fig.scaling}. Note that the
excitonic feature is much stronger than the absorption by onsite
\textit{d-d} transitions in 0D cuprate CuB$_{2}$O$_{4}$
\cite{CuB2O4}, which supports its correlated nature along the
chain.

In summary, we presented carefully measured optical spectra of
Sr$_{2}$CuO$_{3}$. A strong charge transfer peak polarized along
the chain direction shows a strongly asymmetric shape with a clear
gap edge. At low temperature the main charge transfer peak shows a
large blue shift and additional sharp peaks appear. The overall
temperature dependence of the main peak, which is mainly
attributed to the electron phonon coupling, could be scaled by its
peak height and width.  And the additional peaks were attributed
to Wannier-like excitons. Such an excitonic behavior is expected
with the long range Coulomb interaction, which should be important
when $V\sim 2t$ in one-dimensional extended Hubbard model.

\begin{acknowledgments}
K. W. Kim acknowledges discussions with D. Baeriswyl, E.
Jeckelmann, C. Bernhard and A. Dubroka. This work is supported by
the Schweizer Nationalfonds (SNF) with grant 200020- 119784, by
the Department of Energy under contract No. DE-AC02-98CH10886, by
the Creative Research Initiatives (Functionally Integrated Oxide
Heterostructure) of KOSEF, and by the Brain Korea 21 Project in
2002. The experiments at PLS was supported by MOST and POSCO.

\end{acknowledgments}



\begin{references}
\bibitem{EELS1-comment} Note that ref.\cite{EELS1} is often cited as data showing
$V>2t$ which is an unscreened V value. But it is mentioned that
the screened $V$ for 1D EHM is smaller than 2$t$.
\end{references}
\end{document}